\documentclass[aps,prb,twocolumn,superscriptaddress,showpacs,a4paper,preprintnumbers]{revtex4}
\usepackage{graphicx}

\begin{document}
\preprint{LAUR: 03-6352}
\title[Thermodynamic properties of La$_{2}$CuO$_{4.11}$ ]{Thermodynamic properties of excess-oxygen-doped La$_{2}$CuO$_{4.11}$ near a
simultaneous transition to superconductivity and long-range magnetic order}

\author{G. A. Jorge}
\email[Corresponding author: ]{jorge@lanl.gov}
\affiliation{National High Magnetic Field Laboratory, Los Alamos National Laboratory, Los
Alamos, New Mexico 87545}
\affiliation{Departamento de F\'{\i}sica, Universidad de
Buenos Aires, Pabell\'{o}n 1 Ciudad Universitaria (1428) Buenos Aires, Argentina}

\author{M. Jaime}
\affiliation{National High Magnetic Field Laboratory, Los Alamos National Laboratory, Los
Alamos, New Mexico 87545}

\author{L. Civale}
\affiliation{Superconductivity Technology Center, Los Alamos National Laboratory, Los
Alamos, New Mexico 87545}

\author{C. D. Batista}
\affiliation{Center for Non-Linear Studies, Los Alamos National Laboratory, Los Alamos, New
Mexico 87545}

\author{B. L. Zink}
\affiliation{Department of Physics, University of California, San Diego, La Jolla,
California 92093}

\author{F. Hellman}
\affiliation{Department of Physics, University of California, San Diego, La Jolla,
California 92093}

\author{B. Khaykovich}
\affiliation{Department of Physics and Center for Material Science and Engineering,
Massachusetts Institute of Technology, Cambridge, Massachusetts 02139}

\author{M. A. Kastner}
\affiliation{Department of Physics and Center for Material Science and Engineering,
Massachusetts Institute of Technology, Cambridge, Massachusetts 02139}

\author{Y. S. Lee}
\affiliation{Department of Physics and Center for Material Science and Engineering,
Massachusetts Institute of Technology, Cambridge, Massachusetts 02139}

\author{R. J. Birgeneau}
\affiliation{Department of Physics, University of Toronto, Toronto, Ontario M5S 1A7, Canada}

\begin{abstract}
We have measured the specific heat and magnetization {\it versus} temperature
in a single crystal sample of superconducting La$_{2}$CuO$_{4.11}$ and in a
sample of the same material after removing the excess oxygen, in magnetic
fields up to 15~T. Using the deoxygenated sample to subtract the phonon
contribution, we find a broad peak in the specific heat, centered at 50~K.
This excess specific heat is attributed to fluctuations of the Cu spins
possibly enhanced by an interplay with the charge degrees of freedom, and
appears to be independent of magnetic field, up to 15~T. Near the
superconducting transition $T_{c}$($H$=0)=~43~K, we find a sharp feature that
is strongly suppressed when the magnetic field is applied parallel to the
crystallographic $c$-axis. A model for 3D vortex fluctuations is used to scale
magnetization measured at several magnetic fields. When the magnetic field is
applied perpendicular to the $c$-axis, the only observed effect is a slight
shift in the superconducting transition temperature.

\end{abstract}

\date{08/29/03}
%\received[Received]{08/29/03}

%\revised[Revised text]{date}

%\accepted[Accepted text]{date}

%\published[Published text]{date}

\startpage{1}
%\endpage{102}
\maketitle

\section{\label{intro}Introduction}

The high-temperature cuprate superconductors are created by doping of Mott
insulating materials, such as La$_{2}$CuO$_{4}$, which show quasi-2D spin-$\textstyle{1\over 2}$ 
antiferromagnetic (AFM) order. Neutron scattering experiments on La$_{2}$CuO$_{4}$ have shown instantaneous spin correlations in the copper-oxygen
planes that persist to very high temperature because of the very strong
in-plane exchange constant $J$.\cite{kastner-rmp98} Although the 3D AFM order,
resulting from weak interplane coupling, is destroyed by doping of La$_{2}$CuO$_{4}$, the in-plane dynamic correlations survive throughout the
underdoped and superconducting portion of the phase diagram. Therefore, the
interplay between the spin degrees of freedom and superconductivity is an
important part of the current research on high-T$_{c}$ superconductors.

Particulary interesting is the coexistence of static incommensurate spin
density wave (SDW) order with superconductivity in several cuprates (for
recent reviews, see Refs. \onlinecite{kivelson-cond-mat02,sachdev-rmp03}).
Static SDW's have been observed first in La$_{2-x-y}$Nd$_{y}$Sr$_{x}$CuO$_{4}$\cite{tranquada-nature95}
and later in La$_{2-x}$Sr$_{x}$CuO$_{4}$, for $x$
near $\textstyle{1\over 8}$,\cite{kimura-prb99} and La$_{2}$CuO$_{4+y}$.\cite{lee-prb99}
Oxygen-doped La$_{2}$CuO$_{4+y}$ is most interesting in this respect not only
because it has the highest $T_{c}$=~43~K of entire doped La$_{2}$CuO$_{4}$
family of superconductors, but also because the transition to the
superconductivity and the in-plane static long-range magnetic order coincide.
According to neutron scattering and muon-spin rotation experiments, there is a
well-defined phase transition to the magnetically-ordered state at
$T_{m}=T_{c}$.\cite{lee-prb99,savici-prb02} This magnetic order is
2-dimensional with the out-of-plane correlation length of order 3~Cu-O planes.
Above $T_{m}$, in the paramagnetic phase, there is evidence for
long-coherence-length, low-energy spin fluctuations.\cite{lee-prb99} By
applying a magnetic field, which suppresses the superconducting order
parameter, the magnetic order parameter
grows.\cite{katano-prb00,khaykovich-prb02,lake-nature02} This indicates that
there is a competition between the magnetic order and superconductivity, in
spite of the close proximity of $T_{m}$ and $T_{c}$.

To learn more about the superconductivity and magnetism, we have measured the
specific heat, in fields up to 15 T, and the magnetization, in fields up to 7~T,
for La$_{2}$CuO$_{4.11}$. We have removed the excess oxygen from a piece of
a similar oxygen-doped single crystal and have used its specific heat to
estimate the phonon contribution. Subtraction of the latter reveals a broad
peak near 50~K in the specific heat, which is insensitive to magnetic field
and which we associate with spin fluctuations, enhanced by the involvement of
charge degrees of freedom. In addition, there is narrow peak near the
superconducting transition. The height and width of the latter peak, as well
as its suppression by the magnetic field, is consistent with results for a
Sr-doped La$_{2-x}$Sr$_{x}$CuO$_{4+y}$ superconductor of similar hole density.
Analysis of the magnetization indicates strong 3D vortex fluctuations when the
field is applied parallel to the $c$-axis of the sample.

In the next section we discuss the experimental techniques
(sec.~\ref{technique}) and the sample preparation and characterization
(sec.~\ref{sample}). In sec.~\ref{results} we present the results for specific
heat (sec.~\ref{specheat}) and magnetization (sec~\ref{magnetiz})
measurements. In sec~\ref{discussion} we discuss the implication of our
results, and a summary is presented in sec.~\ref{conclusion}.

\section{\label{experimental}Experimental details}

\subsection{\label{technique}Measurement techniques}

The specific heat measurements were made with a thin Silicon Nitride (SiN)
membrane microcalorimeter in fields applied parallel and perpendicular to the
$c$-axis by means of a superconducting magnet up to 15~T. The SiN membrane
microcalorimeters consist in a Silicon frame of 11$\times$11~mm$^{2}$ with a
5$\times$5~mm$^{2}$ SiN membrane, 1~$\mu$m thick. On the front side of the
membrane there are a platinum resistor that works as a heater, and three
different thermometers (one platinum resistor for high temperatures and two
NbSi composite resistors for low temperatures).\cite{delinger-rsi94} On the
back side of the membrane, a gold conduction layer is deposited in order to
thermalize the thermometers and heater, and the sample is attached on it with
a thermally conductive compound. These microcalorimeters have very low heat
capacity addenda, allowing measurements of small samples (in the sub-miligram
range), thin films\cite{delinger-rsi94} or small changes in specific heat of
bulk samples when a small addenda is needed. Recent
measurements\cite{zink-rsi02} have proved that in these devices neither the
membrane thermoconductance nor the addenda change significantly with fields up
to 8~T, whereas our measurements show no change up to 15~T. The film
thermometers on the membrane were calibrated in temperature and magnetic field
against a factory-calibrated Cernox thermometer. The method used to measure
specific heat is the thermal relaxation time technique.\cite{bachmann-rsi72}

Magnetization measurements were performed in a commercial 7~T Quantum Design
SQUID magnetometer, using a scan length of 4~cm (longitudinal magnetization),
with fields applied parallel and perpendicular to the $c$-axis.

\subsection{\label{sample}Sample preparation and characterization}

Single crystals of La$_{2}$CuO$_{4}$ have been grown by the travelling solvent
floating zone technique and subsequently oxidized in an electrochemical cell,
as described previously.\cite{lee-prb99} Thermogravimetric analysis gives an
oxygen excess of 0.11.\cite{lee-prb99,khaykovich-prb02} We study a 3.06~mg
piece of a single crystal of La$_{2}$CuO$_{4.11}$ of dimensions 0.6$\times
$0.6$\times$1.4~mm$^{3}$, where the $c$-axis is along the longer dimension.
This is a small piece of the sample used in previous neutron scattering
experiments.\cite{lee-prb99,khaykovich-prb02}

\begin{figure}
\includegraphics[scale=0.7]{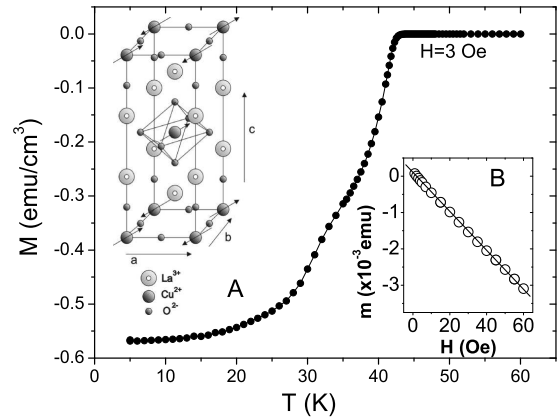}
\caption{Magnetization vs.
temperature at an applied field of 3~Oe. The onset of diamagnetic signal is at
43~K. The structure in the curve at lower temperatures probably indicates a
small amount of remaining stage-6 phase. Inset A: Schematic structure of
La$_{2}$CuO$_{4}$, showing the spins of the copper ions. Inset B: Magnetic
moment vs. magnetic field at low fields (for T=5~K); the line is the result of
a linear fit of the data. Both magnetization measurements were done after
zero-field cooling.}%
\label{fig1}
\end{figure}

The structure of La$_{2}$CuO$_{4}$ is shown in the inset A of Fig.~\ref{fig1}.
The spin-$\textstyle{1\over 2}$~Cu$^{2+}$ ions order antiferromagnetically within the
CuO$_{2}$ planes in the undoped material (as shown in the diagram). The
$c$-axis is perpendicular to the Cu-O planes. In oxygen-doped La$_{2}$CuO$_{4+y}$, the excess oxygen is intercalated between the Cu-O planes and
forms a sine-density-wave, which is periodic along the $c$-axis. This behavior
is called \textit{staging}.\cite{wells-science97} Stage-\textit{n}
corresponds to a period of \textit{n} Cu-O layers. The sample studied here is
predominantly in the stage-4 phase, which has its superconducting transition
at $T_{c}$=~43~K. The onset of static long-range magnetic order coincides with
the onset of superconductivity.\cite{lee-prb99,khaykovich-prb02} Note that
T$_{c}$=~43~K is higher than that of optimally-doped La$_{2-x}$Sr$_{x}$CuO$_{4}$,
which is possibly due to the absence of quenched disorder from a
random distribution of Sr$^{2+}$ dopants. Previous neutron scattering
experiments show no gap in the spin excitations in this
sample.\cite{lee-prb99}

Our sample shows the onset of a diamagnetic signal at 43~K when measuring
magnetization at an applied magnetic field of 3~Oe (see the main panel of
Fig.~\ref{fig1}). The structure observed in the magnetization at around 32~K
almost certainly comes from small inclusions of the stage-6 phase with T$_{c}$=~32~K.

We have estimated the superconducting volume at 5~K from the slope of the
magnetic moment vs. field at low fields (inset B of Fig.~\ref{fig1}),
corrected by a demagnetization factor of $3/2$ for a sphere. The calculated
volume is 0.45~mm$^{3}$, close to the sample volume of 0.43~mm$^{3}$, deduced
from the sample mass. This indicates that the superconductivity is a bulk
phenomenon and develops in the entire sample.

To create the deoxygenated sample we place another 0.5~mg piece of an
oxygen-doped crystal in an Ar flow at 500$^{o}$C for 2 hours. This procedure
eliminates the excess oxygen and results in undoped La$_{2}$CuO$_{4}$
material. We have used this sample to subtract the phonon background from the
specific heat measurements, as discussed below. We note that the specific heat
of the two samples agree very well at low temperatures without any adjustment.

\section{\label{results}Results}

\subsection{\label{specheat}Specific heat experiments}

\begin{figure}
\includegraphics[scale=0.8]{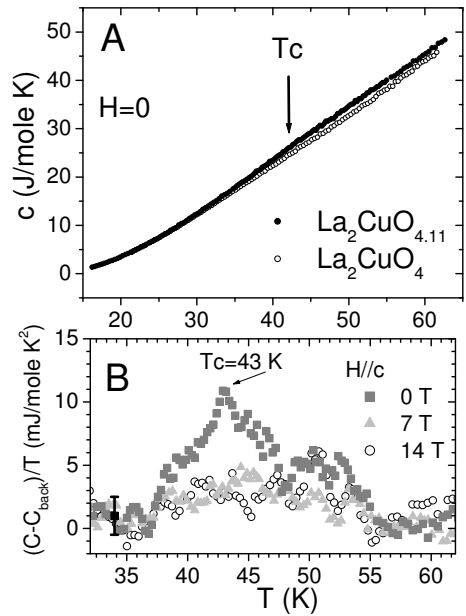}
\caption{Panel A: Specific heat vs.
Temperature at zero field for superconducting La$_{2}$CuO$_{4.11}$ and
insulating La$_{2}$CuO$_{4}$. The arrow indicates the position of the SC
transition in the first curve. Panel B: Specific heat divided by T above the
background at fields of 0, 7 and 14 T applied parallel to the $c$-axis. The
same background, linear in T, is subtracted from all 3 curves. The error bar
shows the typical scattering of points, $\pm$2 mJ/mol K$^{2}$. }%
\label{fig2}
\end{figure}

Fig.~\ref{fig2}A shows the specific heat data for the La$_{2}$CuO$_{4.11}$
single crystal as well as for the deoxygenated crystal of La$_{2}$CuO$_{4}$,
in zero field. The arrow indicates the anomaly at the superconducting
transition for the former. In order to clarify the contribution from
superconductivity, we have subtracted a background (C$_{back}$) linear in T
between 30 and 60~K from the original specific heat data. The measurements
have been made at fields of 0, 7 and 14~T applied parallel to the $c$-axis,
and in Fig.~\ref{fig2}B we plot the quantity (C-C$_{back}$)/T as a function of
temperature. This is the method used perviously to elucidate the specific heat
anomaly at the transition.\cite{migliori-prb90,balbashev-pc96,hirayama-ssc00}
The peak at 43~K at zero field, with full width at half-height of 5~K,
corresponds to the onset of superconductivity. The height of the peak is 10~mJ/mole K$^{2}$,
which is only 2\% of the total specific heat at 43~K. This
peak height is consistent with other measurements of the specific heat in
La$_{2}$CuO$_{4.093}$,\cite{hirayama-ssc00} and in La$_{1.85}$Sr$_{0.15}$CuO$_{4}$.\cite{balbashev-pc96,fisher-prb00} When the field is applied
parallel to the $c$-axis of the sample, the anomaly is suppressed, as
demonstrated by the curves for 7 and 14~T. At these fields, the peak has
broadened and there is no clear sign of an anomaly over the entire temperature
range measured.

\begin{figure}
\includegraphics[scale=0.8]{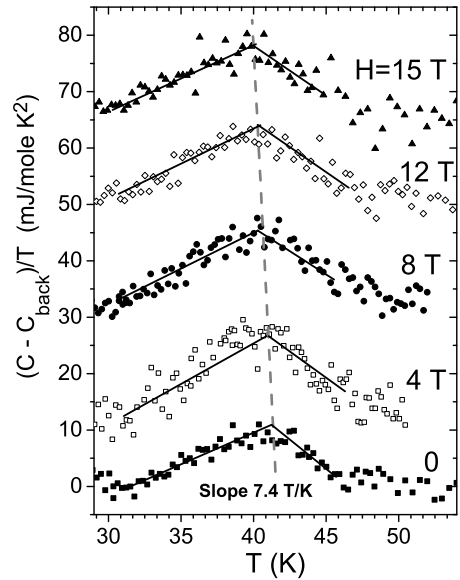}
\caption{Same as Fig.~\ref{fig2}B, with
fields up to 15~T applied in plane. The curves are displaced for clarity, and
straight lines are added as guides to the eye. The dashed line shows the shift
of the SC transition with field.}
\label{fig3}
\end{figure}

In Fig.~\ref{fig3} we show the specific heat data in fields applied
perpendicular to the $c$-axis (in plane). We subtract, as before, a background
linear in temperature between 30 and 60~K and divide by T. The curves are
displaced for clarity. In this case the peak is present in all fields up to 15~T, with no change in height and a small shift towards lower temperatures with
increasing fields (marked with a dashed line on the plot). The rate of change
is 0.13~K/T.

\begin{figure}
\includegraphics[scale=0.8]{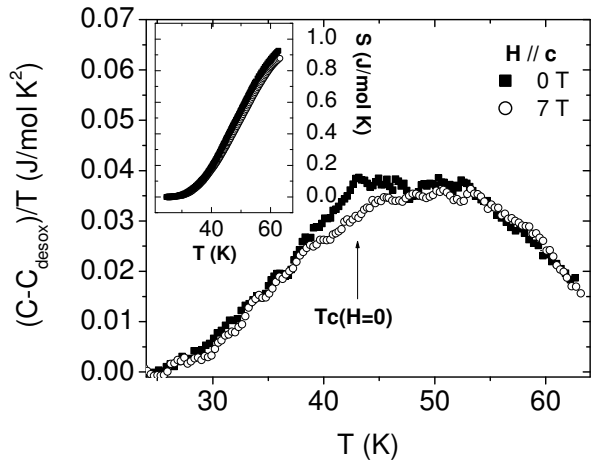}
\caption{Specific heat for La$_{2}%
$CuO$_{4.11}$ with that of the deoxygenated sample subtracted, for zero field
and for a 7~T field applied parallel to the $c$-axis. The superconducting
transition is indicated with an arrow. The inset shows the entropy associated
with the excess specific heat.}
\label{fig4a}
\end{figure}

Although the origin of the sharp peak at zero field in Fig.~\ref{fig2}B and in
fields applied perpendicular to the $c$-axis in Fig.~\ref{fig3} is clearly
identified with the superconducting transition, there is an extra contribution
to the specific heat in the superconducting compound, evident from
Fig.~\ref{fig2}A. The curve for the oxygenated sample departs from the
deoxygenated one above 30~K. Subtracting the specific heat for the
deoxygenated crystal, we obtain a plot of the excess of specific heat, shown
in Fig.~\ref{fig4a}. The curves at the two fields match each other, except for
the superconducting peak in the zero-field curve (indicated with an arrow).
The inset of Fig.~\ref{fig4a} shows the entropy associated with the excess
specific heat obtained by numerical integration. The entropy at 65~K is
0.93~J/mol~K, roughly 20\% of the entropy expected for one spin-$\textstyle{1\over 2}$ per Cu ion
(R~ln~(2S+1)=~5.76~J/mol~K).

\subsection{\label{magnetiz}Magnetization experiments}

\begin{figure}
\includegraphics[scale=0.8]{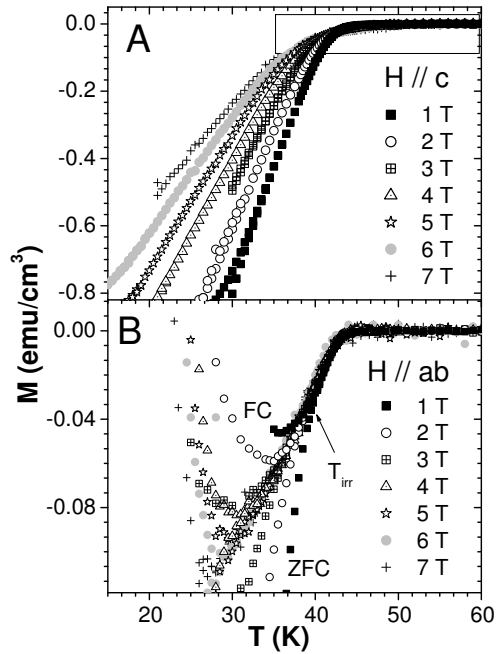}
\caption{Magnetization versus temperature
curves for H parallel (panel A) and perpendicular (panel B) to the $c$-axis.
FC indicates curves taken while field cooling and ZFC curves taken while
zero-field cooling. T$_{irr}$ is the irreversibility temperature (shown for
H=1~T). The rectangle is expanded in Fig.~\ref{fig5}.}
\label{fig4}
\end{figure}

Magnetization measurements have been performed on the same single crystal used
in the heat capacity studies for both orientations of the magnetic field.
Fig.~\ref{fig4} shows the data in fields applied parallel (panel A) and
perpendicular (panel B) to the $c$-axis. In the latter orientation,
irreversible behavior is observed, and the zero field cooled (ZFC, lower
curve) and field cooled (FC, upper curve) magnetization curves merge at the
irreversibility temperature T$_{irr}$, shown on the figure for H=~1~T. In the
reversible region of the magnetization, curves at different fields are
indistinguishable within the scatter of the points. This behavior is
consistent with the specific heat results, which show only a slight
displacement of the superconducting transition temperature with field.

\begin{figure}
\includegraphics[scale=0.7]{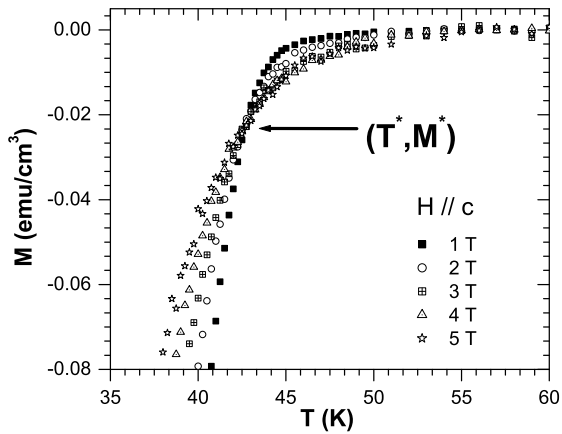}
\caption{Expansion of the rectangular
region indicated in Fig.~\ref{fig4} (H$\parallel$c). The crossing point of
these curves is a distinctive feature of vortex fluctuations. The
characteristic temperature (T$^{\ast}$) and magnetization (M$^{\ast}$) are
indicated.}
\label{fig5}
\end{figure}

For the field applied parallel to the $c$-axis, panel A of Fig.~\ref{fig4}
shows the reversible region of the magnetization up to 7~T. Note that the
scale of the magnetization axis is 10 times than that in panel B. We have
expanded the region marked with a rectangle in Fig.~\ref{fig5}. A clear
crossing point is observed in curves up to 5~T (curves at 6 and 7~T are not
plotted here). The characteristic temperature (T$^{\ast}$=~42.5~K) and
magnetization (M$^{\ast}$=~-0.023~emu/cm$^{3}$) of this crossing point are
marked on the graph. The crossing point in the magnetization vs. temperature
curves is a distinctive feature of vortex fluctuations.
\cite{kadowaki-pc91,kes-prl91,welp-prl91,bulaevskii-prl92,tesanovic-prl92,wahl-prb97,mosqueira-el98,huh-prb02,sutjahja-prb02}

\section{\label{discussion}Discussion}

\subsection{\label{magnet}2D spin fluctuations}

We first discuss the specific heat results. One of the features in the
specific heat of our superconducting sample La$_{2}$CuO$_{4.11}$ is the
absence of a clear magnetic contribution at the simultaneous ordering
temperature of 42-43~K. Indeed, we identified a small anomaly due to the
superconducting transition. However, we did not observe any anomaly associated
with the simultaneous onset of the static incommensurate magnetic order
neither in the absence nor in the presence of the applied field of up to 15~T.
This could be due to one of the following reasons: a)~the entropy recovered at
the magnetic ordering temperature is quite small, as in the parent compound,
\cite{sun-prb91} b)~the entropy is significant but distributed over a broad
temperature range. Therefore, we measured a deoxygenated (insulating) sample
of La$_{2}$CuO$_{4}$ to obtain the phonon contribution to the specific heat.

The insulating sample has two contributions to the specific heat at the low
temperatures studied, one from the phonons and one from magnons. The magnon
contribution (${C_{m}}$) for undoped La$_{2}$CuO$_{4}$ was estimated from
linear spin-wave theory for a 2D square lattice antiferromagnet to
be\cite{sun-prb91}
\begin{equation}
C_{m}=2.385~\mathrm{J/mol}~\mathrm{K}\left(  {\frac{{k_{B}}}{J}}\right)  ^{2}T^{2}
\label{cm}
\end{equation}
where ${k_{B}}$ is the Boltzmann constant. Taking $J$/${k_{B}}$=~1567~K,
eq.~(\ref{cm}) yields ${C_{m}}$=~1.8~mJ/mol~K at 40~K, a value that is four
orders of magnitude less than the phonon contribution at that temperature
(20~J/mol~K), and well below the experimental resolution. We therefore can use
the specific heat of the deoxygenated sample as a measure of the phonon
contribution, leading to the plot of excess specific heat due to the doping
shown in Fig.~\ref{fig4a}.

Doping La$_{2}$CuO$_{4}$ with oxygen leads to three effects which might
contribute to the excess specific heat: (i)~changes in the vibrational
spectrum caused by excess oxygen; (ii)~free carrier excitations; and (iii)~changes
in the spectrum of magnetic excitations. The staging of the
interstitial oxygen is accompanied by a periodic reversal in the direction of
the tilt of the CuO$_{6}$ octahedra along the $c$-axis. This will likely cause
a small change in the phonon spectrum. However, such a difference would give
rise to a change in the specific heat that is monotonic with increasing
temperature, rather than the broad peak we observe. The broad peak is
reminiscent of the Schottky anomaly for a system with a finite number of
excited states, such as a spin-$\textstyle{1\over 2}$ moment in a magnetic field. From the
temperature of the peak we infer that, were such two-level systems responsible
for the excess specific heat, the energy splitting of the two levels would be~$\sim$~7 meV.
We cannot exclude the possibility that the interstitial oxygens
have a two-level systems associated with it, however, we think it is more
likely that the excess specific heat is associated with a disordered component
of the spin system, as discussed below.

The contribution to the specific heat from the free carriers must be small.
Indeed, above $T_{c}$ we use the BCS estimate $\gamma$=$\Delta C/1.43~T_{c}%
$~$\approx$~6~mJ/mol~K$^{2}$, far less than that observed.\cite{tinkham} In
addition, as for any change in the phonon spectrum, free carriers are not
expected to give a peak in the specific heat.

We suggest, instead, that the excess specific heat shown in Fig.~\ref{fig4}
comes from a change in the spectrum of spin excitations resulting from the
doping. The contribution of the spins to the specific heat in the 2D S=$\textstyle{1\over 2}$
square lattice Heisenberg AFM (Quantum Heisenberg Antiferromagnet QHA) become
important at temperatures of order the in-plane exchange constant $J$.
Monte-Carlo simulations of this model show that the heat capacity exhibits a
broad peak at $k_{B}T\simeq0.6~J$
.\cite{gomez-prb89,sutjahja-prb02,sengupta-condmat03} Since the excess
specific heat reported here is observed at 50~K, ascribing the excess specific
heat to that of a QHA would imply that $J$ is suppressed from $\sim135$~meV to
$\sim7$~meV by doping. However, this conclusion contradicts results from
neutron scattering. The instantaneous magnetic correlation length, measured by
integrating over all the energies up to $\sim$~100 meV is almost constant as a
function of temperature in the optimally-doped or slightly underdoped
La$_{2-x}$Sr$_{x}$CuO$_{4}$.\cite{kastner-rmp98} If $J$ were in fact reduced
by doping to 7~meV, then the correlation length would rise dramatically at low
temperatures. Although the instantaneous correlation length has not been
measured in oxygen-doped La$_{2}$CuO$_{4+y}$, its magnetic dynamics is very
similar to slightly underdoped LSCO.\cite{lee-prb99} Furthermore, we have
measured the widths in q-space of the magnetic fluctuation peaks associated
with the SDW order and find no increase in their width up to $\sim9$~meV. This
allows us to place a lower bound on $J$ in the superconductor of
60~meV.\cite{lee-unpublished} We conclude that the excess specific heat is not
described by the QHA with reduced $J$.

Neutron scattering results show that the ordered moment below the SDW
transition increases dramatically with the application of a magnetic
field.\cite{khaykovich-prb02} However, the excess specific heat, aside from
the small component resulting from superconductivity, does not depend on the
applied field. This and the absence of any critical behavior convince us that
the observed anomaly cannot be attributed ${only}$ to the fluctuations of the
ordered magnetic moment near the SDW phase transition. We would not expect
such a contribution since the nearest-neighbor spin correlations are already
established at much higher temperatures, and hence there would not be enough
excess entropy at 50~K to observe experimentally.

There is a connection between the excess specific heat and the
frequency-dependent magnetic susceptibility $\chi^{\prime\prime}(\omega)$
measured with neutron scattering around the $(\pi,\pi)$ point for slightly
underdoped La$_{2-x}$Sr$_{x}$CuO$_{4}$. The latter shows a maximum as the
energy transfer is lowered, at around 7~meV, before going to zero at lower
energy. This maximum is most pronounced near the superconducting transition
temperature.\cite{aeppli-science97} This behavior shows that there is an
enhancement of the $(\pi,\pi)$ spin fluctuations at $\sim7$~meV in La$_{2-x}%
$Sr$_{x}$CuO$_{4}$ of similar hole density to the La$_{2}$CuO$_{4.11}$ sample
studied here. The susceptibility around $(\pi,\pi)$ of La$_{2}$CuO$_{4.11}$
does not show such a maximum, but rather is approximately constant at low
energies. For the latter material $\chi^{\prime\prime}(\omega)$ is enhanced at
low energies relative to that in La$_{2-x}$Sr$_{x}$CuO$_{4}$ because of the
magnons associated with the static long-range-ordered incommensurate SDW.
However, muon measurements indicate that the local magnetic order is fully
developed microscopically at only about 40\% of the muon
sites.\cite{savici-prb02,khaykovich-prb02} The remaining 60\% of the sample,
which has a very small ordered moment, is likely to have spin excitations
closely similar to those in La$_{2-x}$Sr$_{x}$CuO$_{4}$ (x=0.14) and should
exhibit the enhanced spin fluctuations at around 7~meV. We speculate,
therefore, that the broad anomaly in the specific heat reflects the enhanced
magnetic fluctuations coming from the disordered 2D spin-$\textstyle{\frac{1
}{2}}$ system.

Naively, one expects the entropy associated with these fluctuations to be even
smaller than the contribution from the moments that order below 42K.
Consequently, the large enhancement of spin fluctuations observed in the
experiments must the result of an interplay between charge and spin degrees of
freedom. The relevant energy scale for the charge degrees of freedom is given
by the superconducting transition temperature, T$_c$=~43~K, close to the
temperature at which the excess specific heat is observed.

\subsection{\label{SCT}Superconducting transition}

We next discuss the behavior of the magnetization and specific heat near the
superconducting transition. Vortex fluctuations in high temperature
superconductors have been observed in Bi$_{2}$Sr$_{2}$CaCu$_{2}$O$_{8}%
$,\cite{kadowaki-pc91,kes-prl91} YBa$_{2}$Cu$_{3}$O$_{7-\delta}$%
\cite{welp-prl91,wahl-prb97} and La$_{2-x}$Sr$_{x}$CuO$_{4}$
\cite{mosqueira-el98,huh-prb02} among others. Several theoretical studies,
based on the Ginzburg-Landau (GL) free energy, explain the appearance of the
crossing point T$^{\ast}$ where the magnetization M$^{\ast}$ is independent of
the field.\cite{bulaevskii-prl92,tesanovic-prl92} From the values of T$^{\ast
}$ and M$^{\ast}$ we estimate the coherence length along the $c$-axis for the
vortex fluctuations, using the expression in
ref.~\onlinecite{bulaevskii-prl92} and~\onlinecite{tesanovic-prl92}%

\begin{equation}
-M^{\ast}=-M(T^{\ast})=\frac{{k_{B}T^{\ast}}}{{\Phi_{o}l_{c}}}A \label{bula}%
\end{equation}
where k$_{B}$ is the Boltzmann constant, $\Phi$$_{o}$ is the flux quantum,
$l_{c}$ the coherence length for the fluctuations, and the constant $A$ is set
to~1. In our case, the coherence length we find, $\mathit{l}$$_{c}$~$\approx
$125~\AA ,is almost 20 times the distance between planes ($\sim$7~\AA ), which
suggests that the fluctuations in our system have 3D character. For
comparison, the coherence length for La$_{2-x}$Sr$_{x}$CuO$_{4}$ with a Sr
content $x$=0.14 is $\mathit{l}$$_{c}$~$\approx$60~\AA , deducted from the
data in ref.~\onlinecite{huh-prb02}. This value is half the coherence length
in our sample. Also we found that the penetration depth $\lambda$$_{0}%
$=$\lambda$$_{ab}$(T=0) is larger for our material that for the compound doped
with Sr. We estimate the penetration depth for our sample from magnetization
vs. magnetic field curves\cite{bulaevskii-prl92} (that we measured separately
but do not include in this paper) to be $\lambda$$_{0}$$\approx$3800-4000~\AA . This value is in agreement with a previous estimation on powdered
La$_{2}$CuO$_{4+y}$ ($y$$\leq$0.13), which gives $\lambda$$_{0}$$\approx$4200
~\AA .\cite{ansaldo-prb89} For the compound doped with Sr ($x$=0.15), $\lambda
$$_{0}$$\approx$2000-2500~\AA.\cite{kossler-prb87,aeppli-prb87}

Several physical properties show scaling behavior when critical fluctuations
dominate.\cite{ullah-prb91} In particular, magnetization vs. temperature
curves taken at different fields are expected to collapse to a single function
when the variables
\begin{equation}
\frac{M}{(TH)^{s}}\medskip\;\text{vs.\ }\frac{{[T-T}_{c}(H){]}}{{(TH)}^{s}}
\label{variables}%
\end{equation}
are plotted.\cite{wahl-prb97,huh-prb02,welp-prl91} Here the exponent
$s$ is $\textstyle{1\over 2}$ for 2D fluctuations and $\textstyle{2\over 3}$ for 3D behavior. The function
$T_{c}(H)$, the transition temperature as a function of field, is chosen in
order to optimize the scaling.

\begin{figure}
\includegraphics[scale=0.8]{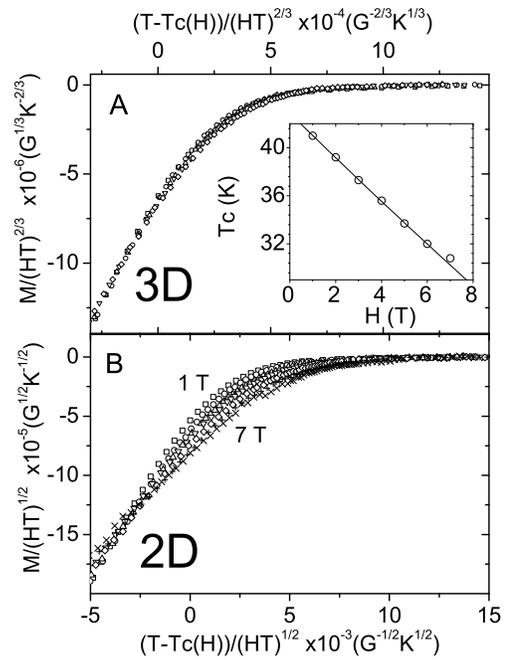}
\caption{Scaling of the magnetization data
in the variables~(\ref{variables}), for 3D (panel A) and 2D (panel B) vortex
fluctuations. The scaling works for presumed 3D fluctuations up to 5~T. Inset:
Function T$_{c}$(H) used in the 3D scaling. All the curves are ZFC.}
\label{fig6}
\end{figure}

Fig.~\ref{fig6} shows the scaling of variables Eq.~(\ref{variables}) for 3D
(panel A) and 2D (panel B) vortex fluctuations, in fields up to 5~T. The
differences between the curves for the 2D case are appreciable, and we are not
able to find a $T_{c}(H)$ function that makes the different curves collapse.
In contrast, panel A shows 3D scaling, for which all the curves up to 5~T
scale within experimental error. The inset in Fig.~\ref{fig6} shows the
function $T_{c}(H)$ used in the scaling. Curves at 6 and 7~T were not included
in Fig.~\ref{fig6}A because they start deviating from the other curves,
presumably due to a departure from the scaling behavior.

Fig.~\ref{fig6} indicates that our system shows 3D vortex fluctuations in
fields applied parallel to the $c$-axis up to 5~T. In higher fields the
fluctuations are still present but their dimensionality becomes obscure,
possibly related to a 3D to 2D crossover at high fields,\cite{mosqueira-el98}
although this point requires more investigation.

As we mentioned earlier, such scaling of magnetization is very common to
different high-T$_{c}$cuprates, including Bi$_{2}$Sr$_{2}$CaCu$_{2}$O$_{8}$,
YBa$_{2}$Cu$_{3}$O$_{7-\delta}$and La$_{2-x}$Sr$_{x}$CuO$_{4}$. We also point
out that the 3D nature of the flux lines serves as another confirmation of the
doping homogeneity of our material. Despite the oxygen-density modulation
associated with staging, based on the results of neutron and NMR measurements
we know that the degree of doping in the Cu-O layers is as homogeneous as in
the best samples of La$_{2-x}$Sr$_{x}$CuO$_{4}$%
.\cite{lee-prb99,khaykovich-prb02} The existence of 3D flux lines with
correlation length of 20~Cu-O layers serves as an additional confirmation of
this conclusion.

The disappearance of the specific heat peak at the superconducting transition
with a magnetic field parallel to the $c$-axis has been universally observed
in cuprate superconductors.
\cite{inderhees-prl91,jeandupeux-prb96,carrington-prb96,junod-pc99,fisher-prb00}
Several mechanisms have been proposed to account for this effect, such as the
possible 2-dimensional nature of the superconducting phase transition in
layered cuprates, effective one-dimensional character of the transition in the
presence of the field because the quasiparticles are bound to the lowest
Landau level in the field, or finite size effects (see, for example,
Ref.~\onlinecite{welp-prl91} and references therein). All these effects are
anisotropic and result in suppression of the transition by the field parallel
to the $c$-axis.

\begin{figure}
\includegraphics[scale=0.8]{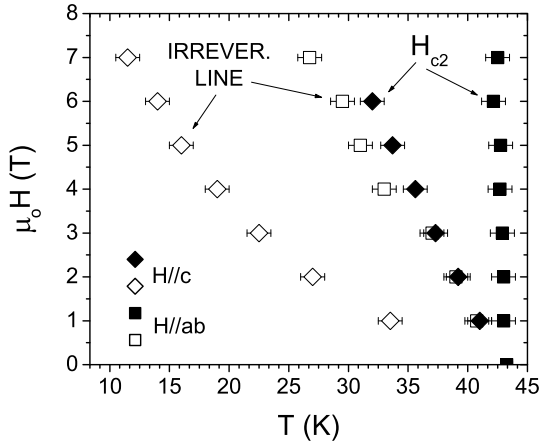}
\caption{Phase diagram from the
magnetization experiment. Open symbols represent irreversibility line, and
solid symbols H$_{c2}$. Squares are for H$\perp$c and diamonds for
H$\parallel$c.}
\label{fig7}
\end{figure}

For completeness, in Fig.~\ref{fig7} we present the phase diagram constructed
from magnetization data. The irreversibility line is taken from the merging of
the ZFC and FC data, the H$_{c2}$line for H$\parallel$c is taken from the
scaling (inset of Fig.~\ref{fig6}), and the H$_{c2}$line for H$\perp$c from
the extrapolation of the reversible magnetization to zero. Note that we have
found very similar values of H$_{c2}(T)$(H $\parallel$c) from the transport
measurements on the same sample.\cite{khaykovich-prb02}

\section{\label{conclusion}Summary}

We have presented specific heat and magnetization measurements on a single
crystal of La$_{2}$CuO$_{4.11}$, in magnetic fields up to 15~T. A broad peak
centered at 50~K is observed in the specific heat measurements when the
specific heat of a deoxygenated sample is subtracted. We attribute this peak
to fluctuations in the disordered spin system associated with the optimally
doped and underdoped superconductors. The effect of these fluctuations is
possibly enhanced by an interplay between the charge and spin degrees of
freedom. At the superconducting transition, we find that a sharp specific heat
peak associated with the onset of the superconducting state is dramatically
broadened, when the field was parallel to the c-axis. Our magnetization
measurements show evidence of strong vortex fluctuations with a correlation
length of 20 Cu-O planes. 3D scaling fits the data well with a quadratic
H$_{c2}$curve up to 5~T.

\begin{acknowledgments}
We thank V. Bekeris for a critical reading of the manuscript. Work at the
NHMFL was performed under the auspices of the National Science Foundation
(DMR90-16241), The State of Florida, and the US Department of Energy. Work at
MIT was supported primarily by the MRSEC Program of the National Science
Foundation under award number DMR 02-13282. Research at the University of
Toronto was supported by the Natural Science and Engineering Research Council
of Canada.
\end{acknowledgments}

\bibliography{guille-boris}

\end{document}